\documentclass{elsart}

\usepackage{epsfig}

\begin{document}

\begin{frontmatter}

\title{The N-Chain Hubbard model in the Composite Operator Method}

\author[a]{A.~Avella}, \author[a]{F.~Mancini\thanksref{Prof}},
\author[a]{M.~M.~S\'anchez\thanksref{Maria}}
and \author[b]{R.~Sridhar}

\address[a]{Universit\`a degli Studi di Salerno -- Unit\`a INFM di Salerno\\
Dipartimento di Scienze Fisiche ``E.~R.~Caianiello''\\84081 Baronissi
Salerno, Italy}
\address[b]{The Institute of Mathematical Sciences\\
C.I.T. Campus, Tharamani, Madras - 600113, India}

\thanks[Prof]{Corresponding author:\\Prof.~ Ferdinando Mancini\\Dipartimento di
Scienze Fisiche ``E.~R.~Caianiello''\\84081 Baronissi (SA), Italy\\FAX:
+39 (0)89 965275\\E-mail: mancini@vaxsa.csied.unisa.it}
\thanks[Maria]{M.~M.~S. acknowledges a grant from the Instituto Nazionale per la Fisica
della Materia (INFM).}

\begin{abstract}
We propose a theoretical framework to describe the ladder systems. The
N-chain Hubbard model has been studied within the Composite Operator
Method. In this scheme of calculations the single-particle Green's
function for any number of coupled chains is obtained by solving
self-consistently a system of integral equations.
\end{abstract}

\begin{keyword}
Ladder Systems. N-chain Hubbard Model. Composite~Operator~Method.
\end{keyword}

\date{29 May 1998}

\end{frontmatter}

\newpage

The study of the so-called ladder materials is an actual hot field in
Condensed Matter Physics. In these systems quantum effects become very
important and lead to a dramatic dependence on the number of coupled
chains that constitute the ladder \cite{dagotrice}. Physical
realizations of such systems are, for instance, vanadyl pyrophosphate
((VO)$_2$P$_2$O$_7$) and the family of layer compounds
Sr$_{n-1}$Cu$_{n+1}$O$_{2n}$. Even-leg ladders have a gap in the spin
excitation spectrum whereas odd-leg ladders present no spin gap
\cite{dagotrice}. Such difference is confirmed by measurements of the
magnetic susceptibility and neutron and muon spin scattering
\cite{experiments}. Ladder systems are studied by different models like
the $t$-$t'$-$J$-$J'$ \cite{dagotto}, Heisenberg \cite{barnes} or
Hubbard \cite{kuroki} models. In this paper we propose a theoretical
framework for these systems. We consider N coupled Hubbard chains and
solve the model in the Composite Operator Method (COM).

We consider a two-dimensional lattice described by the lattice vectors
\begin{equation}
\mathbf{R}_i=\hat xi_x+\hat{y}i_y \hspace{0.5cm} i_x=na\left[-\infty\le n\le
\infty\right] \hspace{0.5cm} i_y = mb \left[1\le m\le N\right]
\end{equation}
For open boundary conditions along the \emph{y}-direction, the Hubbard
model is described by the following Hamiltonian
\begin{equation}
H=\sum_{ij} \left( t_{ij} - \mu\delta_{ij} \right) c^\dagger (i) c(j)
+U\sum_i n_\uparrow(i) n_\downarrow(i)
\end{equation}
where $c(i)$ denotes the electron field at the site $\mathbf{R}_i$ in a
spinor notation; $n_\sigma(i) \equiv c^\dagger
_\sigma(i)c_\sigma(i)$ is the density operator for electrons with spin
$\sigma$; U is the on-site Hubbard interaction and $\mu$ is the chemical
potential. The hopping matrix $t_{ij}$ is given by $t_{ij}
=-2t_x\alpha^x_{i_xj_x} \delta_{i_yj_y} - 2t_y\alpha^y_{i_yj_y}
\delta_{i_xj_x}$ where $t_x$ and $t_y$ are the hopping amplitudes along and between the chains,
respectively. When $N=1$ we put $t_y=0$, in order to recover the
$1$-chain model. $\alpha^x_{i_xj_x}$ and $\alpha^y_{i_yj_y}$ are the
projection operators along the $x$- and $y$-direction, respectively:
\begin{equation}
\alpha^x_{i_xj_x}={1\over 2}\left[\delta_{i_xj_{x+1}}
+\delta_{i_xj_{x-1}}\right] \hspace{0.5cm}
\alpha^y_{i_yj_y}={1\over 2}\left[\hat\delta_{i_yj_{y+1}}
+\hat\delta_{i_yj_{y-1}}\right]
\end{equation}
with
\begin{equation}
\hat\delta _{i_yj_y} =
\left\{
\begin{array}{lll}
0 & \textrm{if} & i_y\ne j_y\\ 0 & \textrm{if} & i_y\
\textrm{and/or}\ j_y=0, N+1\\ 1 & \textrm{if} & i_y=j_y
\end{array}
\right.
\end{equation}
For the prescribed boundary conditions the electron field has the
following representation
\begin{equation}
c(i)={a\over 2\pi}\int _{-\pi/a}^{\pi/a} dk_x e^{ik_xi_x}\sqrt{2\over
N+1}\sum_{k_y} \sin (k_yi_y) c(\mathbf{k})
\end{equation}
\begin{equation}
k_y={\pi l\over b(N+1)} \hspace{0.5cm} 1\le l\le N
\end{equation}
The Fourier transform of $t_{ij}$ takes the expression
\begin{equation}
t_{ij} = {2\over N+1} \sum_{k_y}{a\over 2\pi} \int _{-\pi/a}^{\pi/a}
dk_x e^{i(i_x-j_x)k_x}\sin(k_yi_y)
\sin(k_yj_y)\,t(k_x,k_y)
\end{equation}
where $t(k_x,k_y)=-2t_x\cos(k_xa)-2t_y\cos(k_yb)$; $a$ and $b$ are the
lattice constants along the $x$ and $y$ direction, respectively. In the
framework of the \emph{COM} \cite{COM} we consider the Hubbard doublet
$\psi^\dagger(i)=(c^\dagger(i)(1-n(i)),c^\dagger(i)n(i))$ as the basic
field, where $n(i)=c^\dagger(i) c(i)$ is the density operator. By means
of the Hamiltonian (2) the Heisenberg equation for the composite field
is given by
\begin{equation}
i{\partial\over \partial t} \psi(i) = j(i) = \left(\begin{array}{c}
-\mu\xi(i) +c^t(i)+\pi(i) \\ -(\mu-U)\eta(i)-\pi(i)\end{array}\right)
\end{equation}
with $c^t(i)\equiv \sum_j t_{ij} c(j), n_\mu(i) = c^\dagger(i) \cdot
\sigma_\mu c(i)$,
$\pi(i)={1\over 2}\sigma^\mu n_\mu (i) c^t(i)+c(i)
[c^{t^\dagger}(i)\cdot c(i)]$.

Let us consider the thermal retarded Green's function $S(i,j)=\langle
R[\psi(i)\psi^\dagger(j)]\rangle$. In the two-pole approximation
\cite{COM} we have the equation
\begin{equation}
\left[ i{\partial\over \partial t} \delta_{il} - \epsilon (i,l) \right]
S(l,j) = i\delta (t_i-t_j)\,I(i,j)
\label{eq12}
\end{equation}
where $I(i,j) = \left \langle \left\{ \psi(i), \psi^\dagger (j)\right\}
\right \rangle$ and $m(i,j) = \left\langle \left\{ i{\partial\over \partial t} \psi(i),
\psi^\dagger(j) \right\} \right\rangle = \epsilon(i,l)\,I(l,j)$.

By introducing the Fourier transform
\begin{eqnarray}
S(i,j) &=& {ia\over (2\pi)^2} \int_{-\infty}^{+\infty} d\omega\,
e^{-i\omega(t_i-t_j)} \int^{\pi/a}_{-\pi/a} dk_x\,
e^{ik_x(i_x-j_x)}\nonumber\\
&&{2\over N+1} \sum_{k_y}{2\over N+1}\sum _{p_y} \sin (k_yi_y)\sin
(p_yj_y)\,S(k_x,k_y, p_y, \omega)
\end{eqnarray}
Eq.~(\ref{eq12}) takes the following form in momentum space
\begin{equation}
\omega S (k_x,k_y, p_y, \omega)+ {2\over N+1}\sum _{q_y} \epsilon(k_x,
k_y, q_y) S (k_x,q_y, p_y, \omega)
=I(k_y, p_y)
\end{equation}

The solution of this integral equation will determine the Green's
function, once the matrices $I(k_y,p_y)$ and $\epsilon(k_x,k_y, q_y)$
are known. Lengthy but straightforward calculations show that the
normalization matrix $I(k_y,p_y)$ and energy matrix $\epsilon(k_x,k_y, q_y)$
are determined by
\begin{equation}
I(k_y,p_y) = \sum_{i_y}\sin (k_yi_y \sin (p_yi_y) I(i_y)
\end{equation}
\begin{equation}
\epsilon(k_x,k_y, q_y)={2\over N+1}\sum _{q_y} m(k_x,k_y, q_y)
\sum_{i_y}\sin(q_yi_y) \sin(p_yi_y) I^{-1}(i_y)
\end{equation}
where
\begin{equation}
I(i_y) = \left( \begin{array}{cc}
1-{1\over 2}\langle n(i_y)\rangle & 0\\ 0 & {1\over 2}\langle
n(i_y)\rangle\end{array} \right)
\end{equation}
The elements of the $m$-matrix are given by
\begin{eqnarray}
m_{11}(k_x, k_y, p_y) &=& -\mu I_{11}(k_y, p_y) +\Delta(k_y, p_y) +
t(k_x, k_y) \left[I_{11}(k_y, p_y) \right.\nonumber \\
&& \left.- I_{22}(k_y, p_y)\right] + P(k_x, k_y, p_y) \nonumber \\
m_{12}(k_x, k_y, p_y) &=& -\Delta(k_y, p_y) + t(k_x, k_y) - P(k_x, k_y,
p_y)\\
m_{22}(k_x, k_y, p_y) &=& -(\mu-U) I_{22}(k_y, p_y) +\Delta(k_y,
p_y) + P(k_x, k_y, p_y)\nonumber
\end{eqnarray}
with the following definitions
\begin{eqnarray}
&&\Delta(k_y, p_y) = \sum_{i_y}\sin(k_yi_y)
\sin(p_yi_y)\left[\langle\xi^t(i)\xi^\dagger (i)\rangle -
\eta^t(i)\eta^\dagger (i)\rangle\right] \\       %
(21)
&&P(k_x, k_y, p_y) = -t_x\cos(k_xa)\left[ P^{x+}(k_y, p_y) + P^{x-}
(k_y, p_y)\right] -\nonumber \\
&& \qquad -t_y\left[ P^{y+}(k_y, p_y) + P^{y-} (k_y, p_y)\right]
\nonumber\\
&& P^{x\pm}(k_y, p_y) = \sum_{i_x} \sin (k_yi_y)\sin (p_yi_y) P^{x\pm}
(i_y) \\    %            (22)
&& P^{y\pm}(k_y, p_y) = \sum_{i_x} \sin (k_yi_y)\sin (p_yi_y\mp p_yb)
P^{y\pm} (i_y) \nonumber \\
&& P^{x\pm}(i_y) = {1\over 4} \langle n_\mu (i) n_\mu (i_x \mp a,
i_y)\rangle - \langle c_\uparrow(i)
c_\downarrow (i)c^\dagger_\downarrow (i_x\mp a, i_y) c^\dagger_\uparrow
(i_x \mp a, i_y)\rangle \nonumber \\
&& P^{y\pm}(i_y) = {1\over 4} \langle n_\mu (i) n_\mu (i_x, i_y \mp b,
i_y)\rangle \nonumber \\
&&\qquad - \langle c_\uparrow(i)
c_\downarrow (i)c^\dagger_\downarrow (i_x, i_y\mp a, i_y)
c^\dagger_\uparrow (i_x,i_y \mp b, i_y)\rangle  %     (23)
\end{eqnarray}
As we see a series of unknown parameters appear. This is a consequence 
of the fact that the 
properties of the composite fields are not known  \emph{a priori}. They are
determined by the dynamics and the boundary conditions and must 
be self-consistently calculated. Parameters such 
as $\mu,\langle n(i_y)\rangle$ and $\Delta(i_y)$ can be expressed in
terms of matrix elements of the Green's
function. The parameters $P^x(i_y)$ and $P^{y\pm} (i_y)$ are static
spin, charge and pair correlation
functions and can be determined by the content of the Pauli principle. 
We thus can write a 
series of coupled  equations which lead to a self-consistent calculation
of the Green's function.

\begin{figure}[tb]
\begin{center}
\epsfig{file=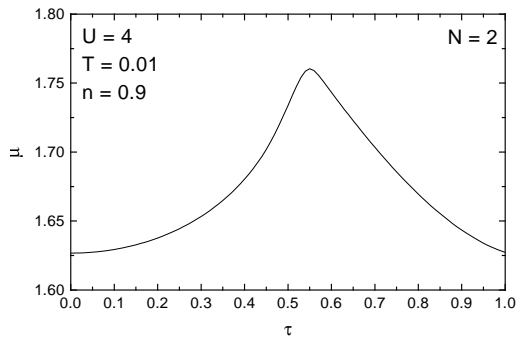,width=7.5cm,clip=}
\end{center}
\caption{Chemical potential $\mu$ as a function of the ratio $\tau = t_y/t_x$
for $U = 4$, $T = 0.01$ and $n = 0.9$ in the two-leg ladder ($N = 2$).}
\label{fig1}
\end{figure}

As preliminary result, the chemical potential as a function of the ratio
$\tau = t_y/t_x$ for the two-leg ladder system ($N = 2$) is presented in
Fig.~\ref{fig1}. The sharp feature around $\tau \approx 0.55$ could
indicate the onset of some spin ordering related to the increasing of
frustration in the spin coupling channel due to the appearance of an
additional exchange interaction along the rungs. A deeper comprehension
of the on-going dynamics requires the analysis of the correlation
functions of the system.

\end{document}